\documentclass[twocolumn]{aastex63}

\usepackage{amssymb, amsfonts, amsmath,framed}
\usepackage{comment}
\usepackage{enumitem}

\usepackage{bm}
\expandafter\ifx\csname package@font\endcsname\relax\else
 \expandafter\expandafter
 \expandafter\usepackage
 \expandafter\expandafter
 \expandafter{\csname package@font\endcsname}
\fi
\expandafter\ifx\csname package@font\endcsname\relax\else
 \expandafter\expandafter
 \expandafter\usepackage
 \expandafter\expandafter
 \expandafter{\csname package@font\endcsname}
\fi
\def\bi{\begin{itemize}}
\def\ei{\end{itemize}}
\def\bq{\begin{equation}}
\def\eq{\end{equation}}
\def\bqy{\begin{eqnarray}}
\def\eqy{\end{eqnarray}}

\begin{document}
\title{Magnetic Field-Line Curvature and Its Role in Particle Acceleration by Magnetically Dominated Turbulence} 

\author{Samuel T. Sebastian}
\affiliation{Department of Physics, University of Florida, Gainesville, FL 32611, USA}
\email{sebastian.s@ufl.edu}
\author{Luca Comisso}
\affiliation{Department of Physics, Columbia University, New York, NY 10027, USA}
\affiliation{Department of Astronomy, Columbia University, New York, NY 10027, USA}
\email{luca.comisso@columbia.edu}

\begin{abstract}
We employ first-principles, fully kinetic particle-in-cell simulations to investigate magnetic field-line curvature in magnetically dominated turbulent plasmas and its role in particle acceleration through curvature-drift motion along the motional electric field. By varying the fluctuation-to-mean magnetic-field ratio $\delta B_0/B_0$, we examine curvature $\kappa$ statistics and their connection to particle acceleration. The curvature probability densities display broad power-law wings, scaling linearly in $\kappa$ below the peak and developing hard high-$\kappa$ tails for $\delta B_0/B_0 \gtrsim 1$. As the mean field strengthens, the high-$\kappa$ tails steepen, and large-curvature events are suppressed when $\delta B_0/B_0 \ll 1$. The probability density functions of magnetic field-line contraction, ${\bm v}_E \cdot {\bm\kappa}$, with ${\bm v}_E$ the field-line velocity, develop power-law tails well described by a symmetric Pareto distribution, characteristic of stochastic and intermittent energy exchanges, with the tails becoming harder as $\delta B_0/B_0$ increases. Our guiding-center analysis shows that curvature-drift acceleration accounts for a substantial fraction of the energization via the motional electric field, and that it strengthens with increasing $\delta B_0/B_0$. For well-magnetized particles, curvature-drift acceleration typically exceeds ${\bm\nabla}B$ drift, polarization drift, and betatron contributions. These results identify curvature-drift acceleration as a principal pathway through which magnetized turbulence transfers energy to nonthermal particles in astrophysical plasmas.

\vspace{0.9cm}

\end{abstract}

\section{Introduction} \label{sec:intro} 

Energetic nonthermal particles are a common feature of high-energy astrophysical plasmas. Observations across diverse sources indicate that a small fraction of the plasma population can acquire a disproportionately large share of the energy, often forming extended power-law distributions \citep{Longair2011,GER2016}. These energetic particles are responsible for much of the high-energy radiation and secondary messengers observed in the universe, and their properties encode essential information about the plasma conditions, magnetic field structure, and energy dissipation processes \citep{RybLig79,DermerMenon2009}. Understanding how they are accelerated is therefore central to interpreting observations and modeling astrophysical environments. 

Magnetized turbulence offers a natural pathway for such acceleration and has been invoked to explain diverse phenomena, including the hard radio spectra of pulsar wind nebulae \citep[e.g.][]{Lyutikov19,Comisso20ApJL}, the rapid gamma-ray variability of active galactic nuclei jets \citep[e.g.][]{Narayan12,Sobacchi23}, neutrino production from accelerated protons in black hole coronae \citep[e.g.][]{Murase20,Fiorillo24}, the prompt emission of gamma-ray bursts \citep[e.g.][]{Thompson06,Asano15}, and the acceleration of ultra-high-energy cosmic rays \citep[e.g.][]{cfm24,Wang24}. In such systems, the magnetization parameter $\sigma$—the ratio of magnetic energy density to plasma enthalpy density—can exceed unity, implying that the Alfv{\'e}n wave speed approaches the speed of light. This regime, commonly referred to as the relativistic turbulence regime, provides especially favorable conditions for rapid particle acceleration.

First-principles (particle-in-cell, PIC) simulations have demonstrated that magnetically dominated ($\sigma > 1$) turbulence produces a nonthermal particle population with a power-law energy distribution, $dN/d\gamma \propto \gamma^{-p}$, independent of microscopic kinetic scales \citep{Comisso18,Comisso19} (for results on moderately magnetized turbulence, see \citet{Zhd18}). The power-law index $p$ of the nonthermal tail was shown to depend strongly on $\delta B_0/B_0$ \citep{Comisso18, Comisso19,Comisso20ApJL}, the ratio of fluctuation to mean magnetic field, with decreasing (increasing) $\delta B_0/B_0$ leading to an increase (decrease) of $p$. The power-law index was found to be regulated by non-resonant stochastic acceleration driven by the motional electric field, which results in a diffusion coefficient in energy space $D_\gamma(\gamma) = \kappa_{\rm acc} \sigma_0 (\delta B_0/B_0)^{2} \gamma^2 c/{l_c}$ \citep{Comisso19}, where $l_c$ indicates the turbulence coherence length and $\kappa_{\rm acc} \simeq 0.1$ from PIC simulation measurements. Consequently, the mean acceleration time $t_{\rm acc} = \gamma^2/4 D_\gamma$ decreases steeply with increasing fluctuation-to-mean field ratio. 

In this work, we focus on how particles gain energy from the motional electric field induced by turbulent plasma motions, with particular attention to acceleration via curvature-drift motion along the motional electric field. PIC simulations suggest that shear of the motional electric field along magnetic field lines is important for reconstructing particle energy histories \citep{Bresci22}, and analytical frameworks have discussed in detail the different forces acting on the particles \citep{Lem21,Lem22,Lem25}. Here, we employ a guiding-center framework to provide a quantitative assessment of curvature-drift acceleration and its dependence on $\delta B_0/B_0$. Guiding-center studies have extensively explored curvature-drift acceleration in magnetic reconnection \citep{Drake06,Dahlin14,Guo14,Guo15,LiX15,LiX17,Dahlin17,Rowan19}, where ideal electric fields are shaped by large-scale reconnection outflows. In turbulence, motional electric fields are intermittent and disordered, and the tangled magnetic geometry creates numerous potential sites for curvature-drift acceleration. Understanding this process is key to explaining how turbulence transfers energy to nonthermal particles and to linking kinetic-scale plasma physics with observable astrophysical signatures.

\section{Numerical Method} \label{sec:model} 

We perform fully kinetic simulations by solving the Vlasov-Maxwell system of equations with the particle-in-cell method \citep{birdsall85}, using the code TRISTAN-MP \citep{buneman93,spitkovsky05}. 
The computational domain is a cubic box of size $L^3$ with periodic boundaries, discretized into $750^3$ uniform cells. 

The plasma consists of electrons and a single ion species (protons) with charge number $Z=1$ and mass $m_i = 1836 \, m_e$. The combined number density is $n_0 = n_{e0} + n_{i0}$. Particle velocities are sampled from a Maxwell-J\"{u}ttner distribution with equal initial temperatures $T_0$ for electrons and ions. A uniform background magnetic field ${\bm{B}}_0  = B_0 {\bm{\hat e}}_z$ is imposed along the $z$ axis. Turbulence is seeded by initializing a spectrum of magnetic fluctuations polarized transverse to ${\bm{B}}_0$ (see \citet{Comisso18,Comisso19,Comisso21,Nattila2021} for details), with a root-mean-square amplitude $\delta B_0 = \langle \delta B^2 \rangle^{1/2}$ and an outer coherence length $l_c \simeq L/3$. 

The strength of the fluctuations is parameterized by the magnetization parameter $\sigma_{\delta B}= {\delta B_0^2}/{4\pi h_0}$, where $h_0$ is the enthalpy density, accounting for both ion and electron contributions. Since we are interested in the magnetically dominated regime, we take $\sigma_{\delta B} = 10$, which yields an Alfv{\'e}n speed associated with the fluctuating field ${v_{A}} = c [{\sigma_{\delta B}}/(1 + {\sigma_{\delta B}})]^{1/2} \simeq c$. The ion temperature is initialized to $k_B T_0 = 0.1 m_i c^2$, yielding $\gamma_{{\rm th},i} \simeq 1.167$. Our findings are insensitive to the initial thermal spread, which affects only an overall energy rescaling \citep{Comisso19}.

We adopt grid cells of size $\Delta x = \Delta y = \Delta z = d_{e0}/3$, where $d_{e}=c/\omega_{pe}$ is the initial electron inertial length, with $\omega_{pe} = (4\pi n_0 {e^2}/\gamma_{{\rm th},e} m_e)^{1/2}$ the electron plasma frequency and $\gamma_{{\rm th},e}$ the electron thermal Lorentz factor. The ion inertial length is $d_i=c/\omega_{pi}$, with $\omega_{pi} = ({4\pi n_0 {e^2}/{\gamma_{{\rm th},i} m_i}})^{1/2}$ the ion plasma frequency and $\gamma_{{\rm th},i}$ the ion thermal Lorentz factor. Expressed in terms of the ion inertial length, the simulation domain has size $L/d_i = 125$. We use an average of $20$ computational particles per cell, which adequately resolves the nonthermal particle acceleration process, as shown in \citet{Comisso18,Comisso19}. The time step is chosen according to the Courant–Friedrichs–Lewy condition, and simulations are run for $t \geq 10 l_c/c$. We explore three values of the fluctuation-to-mean magnetic field ratio by varying $B_0$, such that $\delta B_0/B_0 \in \left\{ {1/3,1,3} \right\}$. The case $\delta B_0/B_0=1$ serves as the reference simulation. This choice provides a straightforward means to assess how the mean magnetic field influences the statistics of field-line curvature and particle acceleration.

\section{Turbulence Statistics} \label{sec:turbstat} 

\begin{figure}
\begin{center}
   \includegraphics[width=8.65cm]{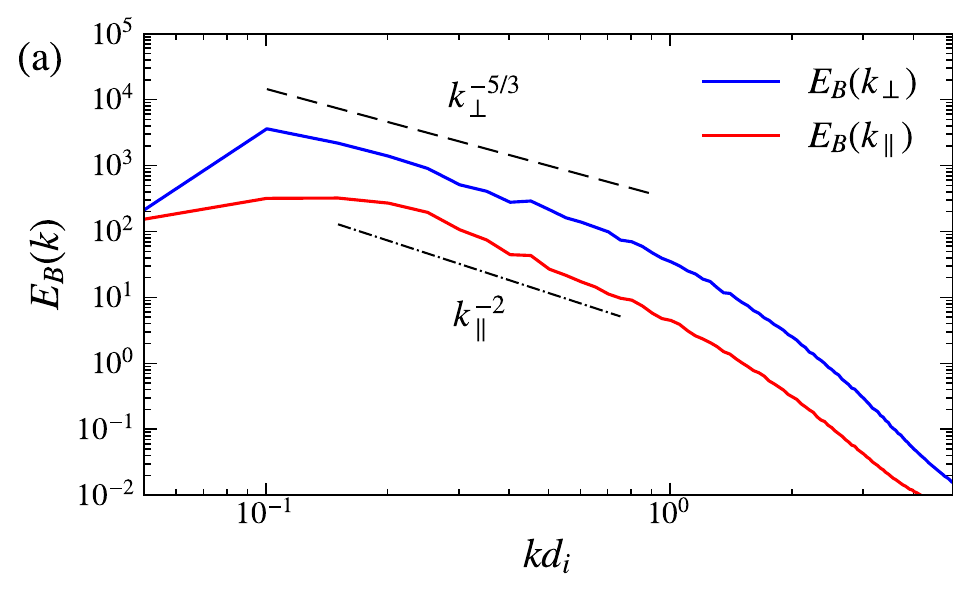}
   \includegraphics[width=8.65cm]{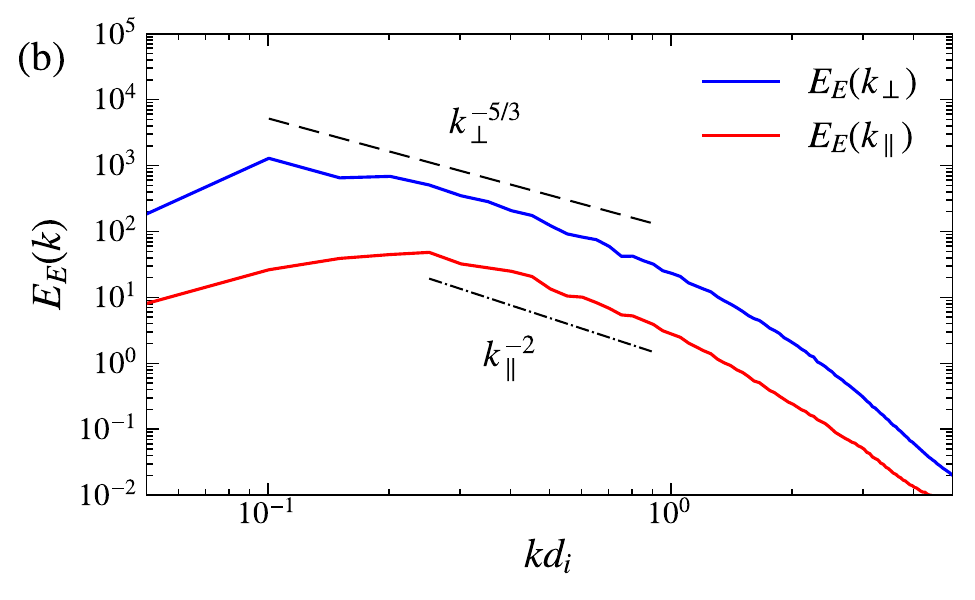}
\end{center}
\vspace{-0.5cm}
   \caption{Perpendicular and parallel spectra of (a) magnetic and (b) electric fluctuations, from the simulation with $\delta B_0/B_0=1$ at $t=2\,l_c/c$. Power-law slopes of $k_\perp^{-5/3}$ and $k_\parallel^{-2}$ are shown for reference. Here, perpendicular ($\perp$) and parallel ($\parallel$) are measured with respect to ${\bm{B}}_0$.}
\label{fig1}
\end{figure}

In our simulations, turbulence develops from the initialized large-scale magnetic fluctuations and subsequently freely decays as no continuous driving is imposed. A fully developed turbulent cascade is established on the outer-scale nonlinear time $l_c/c$. Figure~\ref{fig1} shows the energy spectra of magnetic-field (a) and electric-field (b) fluctuations, for the simulation with $\delta B_0/B_0=1$ at $t = 2 \, l_c/c$.\footnote{We perform most of our analysis at this early time because the particle acceleration rate decreases as the turbulence decays and magnetic fluctuations weaken (see \citet{Comisso18,Comisso19} for its time evolution).} 
We compute the spectra as functions of wavenumbers perpendicular ($k_\perp$) and parallel ($k_\parallel$) to the mean magnetic field ${\bm{B}}_0$.
In the inertial range, $k_{\perp,\parallel} \lesssim d_{i}^{-1}$, the perpendicular spectral slopes are close to $-5/3$, consistent with the prediction for an anisotropic, critically balanced cascade of Alfv{\'e}n waves \citep{GS95,TB98}, while the parallel spectra are slightly steeper, with a slope of roughly $-2$ over a limited range. At scales approaching $d_i$, both perpendicular and parallel spectra steepen, as kinetic physics begins to govern the cascade dynamics \citep[e.g.,][]{Chen14}.

\begin{figure}
\begin{center}
   \includegraphics[width=8.65cm]{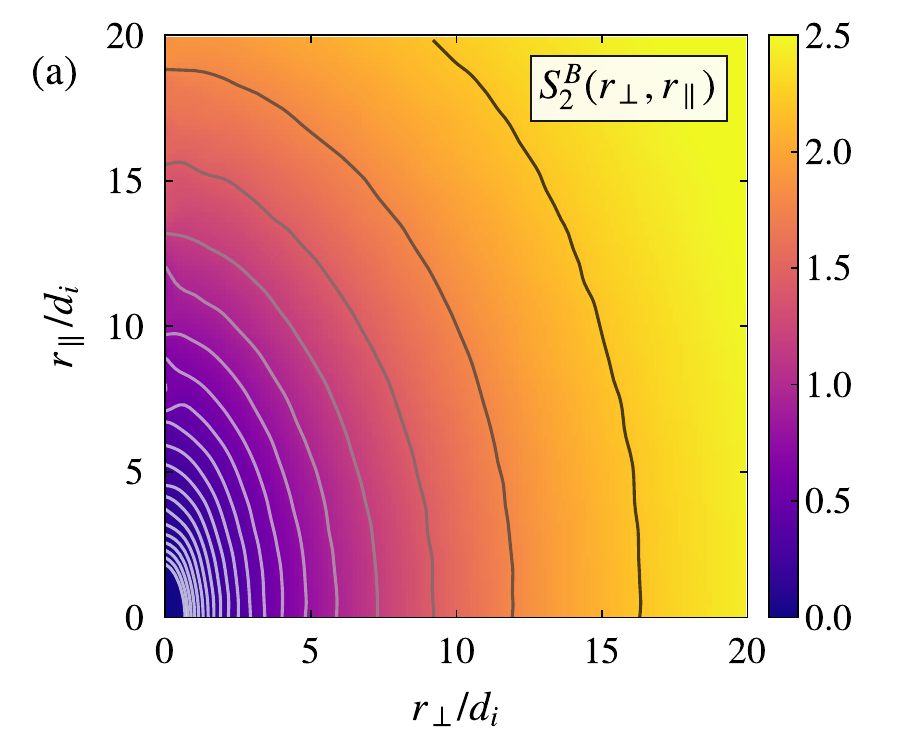}
   \includegraphics[width=8.65cm]{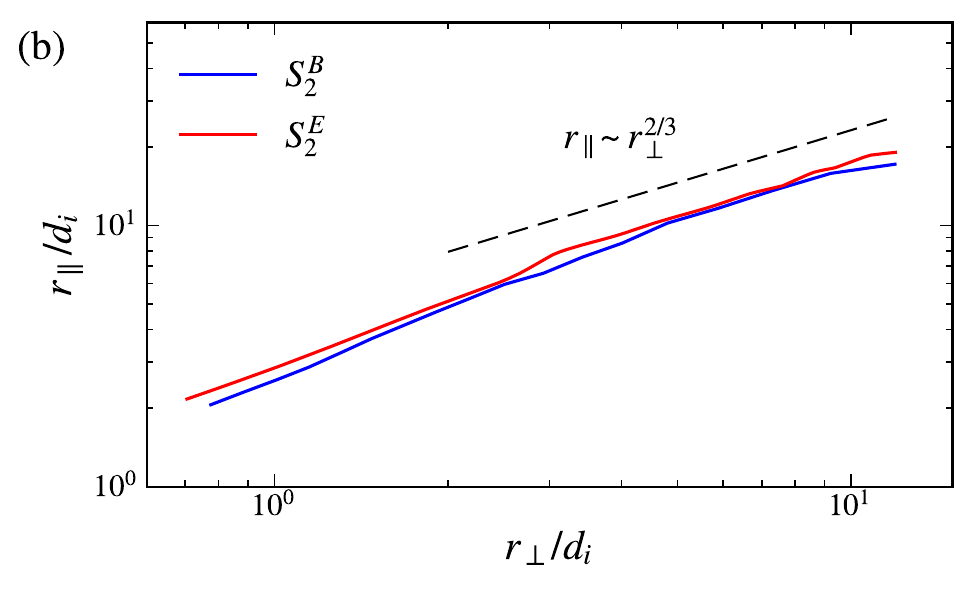}
\end{center}
\vspace{-0.5cm}
   \caption{Second-order structure function of magnetic fluctuations (a) and anisotropy of magnetic and electric fluctuations (b), from the simulation with $\delta B_0/B_0=1$ at $t=2\,l_c/c$. Here, perpendicular ($\perp$) and parallel ($\parallel$) are measured with respect to the scale-dependent local mean magnetic field, as detailed in \citet{CV2000}.}
\label{fig2}
\end{figure}

\begin{figure}
\begin{center}
   \includegraphics[width=8.65cm]{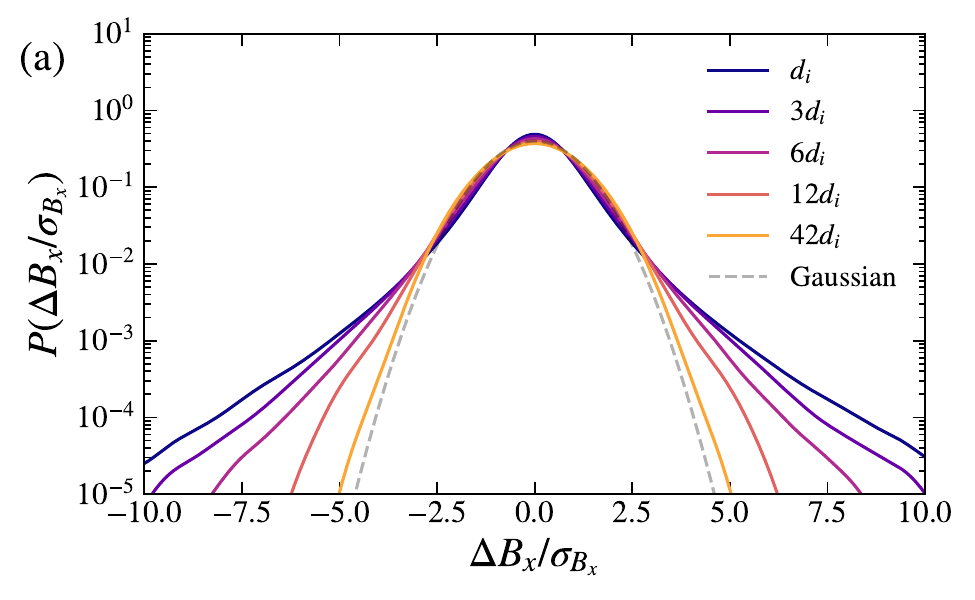}
   \includegraphics[width=8.65cm]{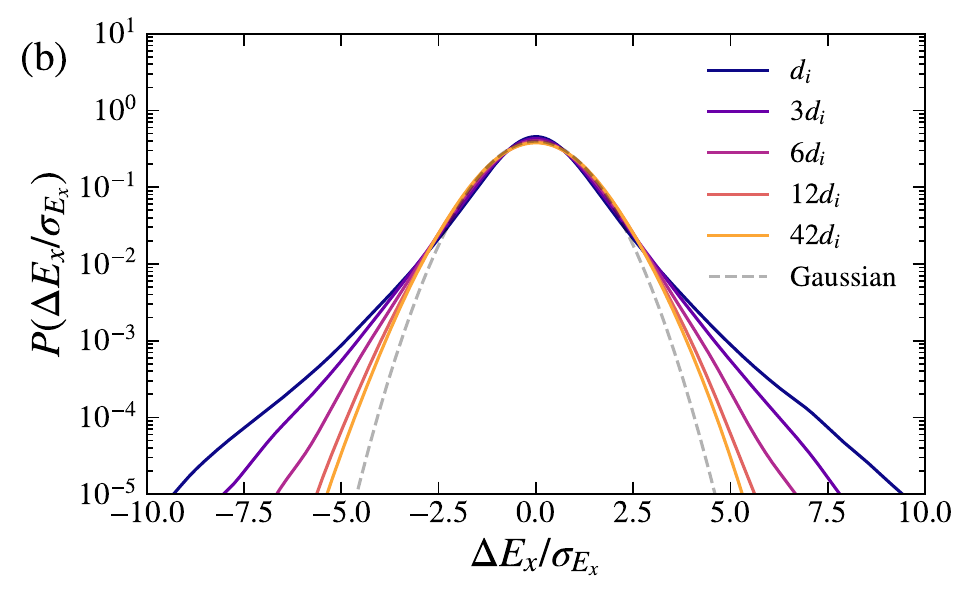}
\end{center}
\vspace{-0.5cm}
   \caption{Probability density functions of field increments at different spatial separations, from the outer scale $r = 42d_i$ down to the ion inertial scale $r = d_i$, including intermediate scales. Results are from the reference simulation ($\delta B_0/B_0=1$) at $t=2\,l_c/c$. The increments $\Delta f(r)$ are (a) $\Delta B_x(r)$ and (b) $\Delta E_x(r)$, each normalized by their standard deviation at the corresponding $r$. At small separations, the PDFs show pronounced non-Gaussian heavy tails, while at large $r$ they approach a Gaussian distribution (dashed curve).}
\label{fig3}
\end{figure}

To characterize the anisotropy of turbulent fluctuations with respect to the local magnetic field, we compute two-point, second-order structure functions, $S_2^Q(r_\parallel,r_\perp) = \langle \left| {\bm{Q}}({\bm{x}} + {\bm{r}}) - {\bm{Q}}({\bm{x}}) \right|^2 \rangle$, for ${\bm Q}={\bm B},{\bm E}$. The separations $r_\parallel$ and $r_\perp$ are defined relative to the scale-dependent local mean magnetic field, following the procedure of \citet{CV2000}. Figure~\ref{fig2}(a) shows the magnetic-field structure function, $S_2^B$, whose isocontours indicate elongated eddies with $r_\parallel \gg r_\perp$ and increasing anisotropy at smaller $r_\perp$. The same qualitative behavior is observed for $S_2^E$. This scale-dependent anisotropy is quantified in Fig.~\ref{fig2}(b) by fitting the major and minor axes of the isocontours. The fits yield $r_\parallel \propto r_\perp^{2/3}$ in the inertial range, consistent with the Goldreich–Sridhar critical balance scaling \citep{GS95}. 
At scales approaching $d_i$, the slope steepens, reflecting the transition to the kinetic turbulence regime. 

\begin{figure}
\begin{center}
   \includegraphics[width=8.65cm]{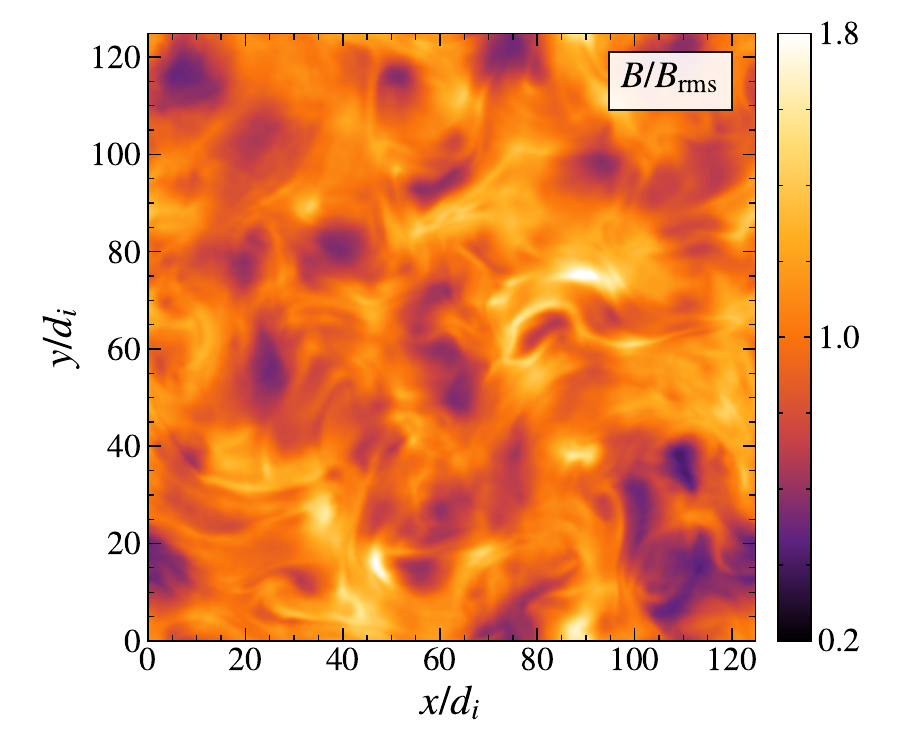}
   \includegraphics[width=8.65cm]{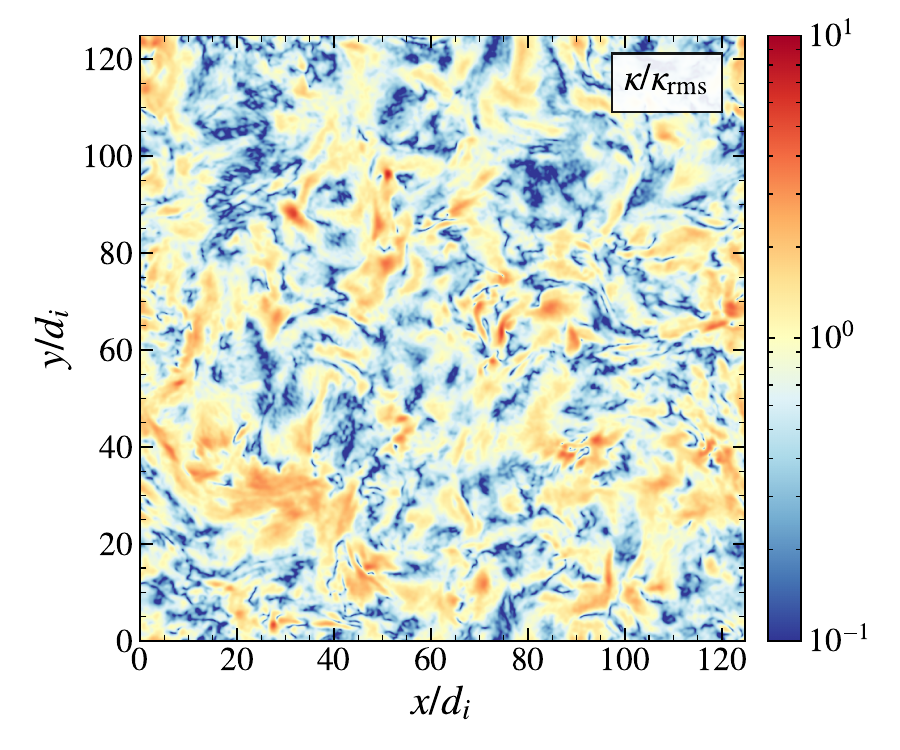}
   \includegraphics[width=8.65cm]{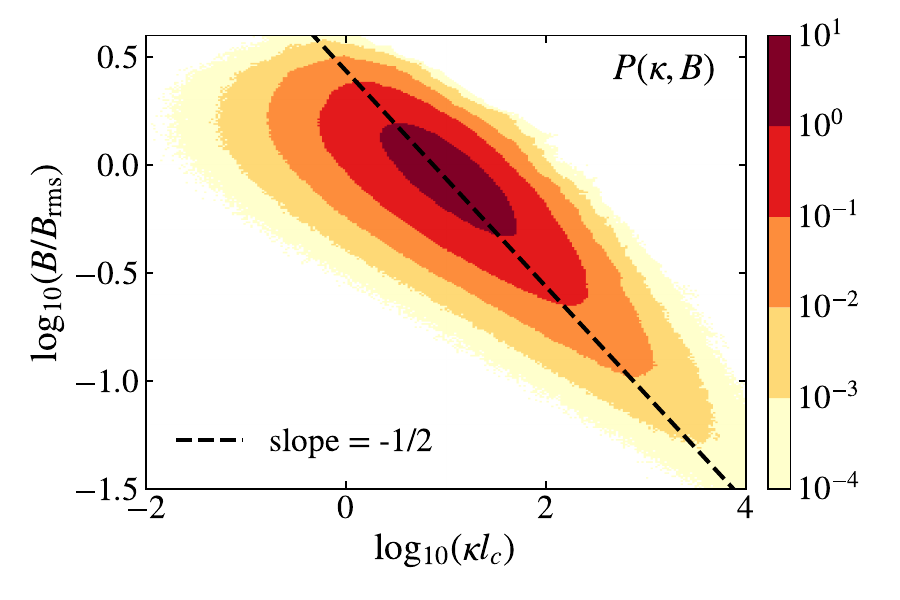}
\end{center}
\vspace{-0.5cm}
   \caption{Two-dimensional slices in the $x$–$y$ plane at fixed $z$ from the reference simulation ($\delta B_0/B_0=1$) at $t=2\,l_c/c$, showing the normalized magnetic field strength $B/B_{\rm rms}$ and the normalized magnetic field-line curvature magnitude $\kappa/\kappa_{\rm rms}$. The bottom panel shows the joint distribution of $B/B_{\rm rms}$ and $\kappa$ (measured in units of the inverse coherence length, $l_c^{-1}$), showing that regions of strong curvature tend to coincide with weaker magnetic fields. A power-law slope of $-1/2$ is provided for reference.}
\label{fig4}
\end{figure}

Beyond anisotropy, relativistic magnetized turbulence exhibits strong intermittency \citep{Davis24}. We quantify this through the scale dependence of probability density functions (PDFs) of field increments $\Delta f(r)$, which increasingly deviate from Gaussianity as the spatial separation $r$ decreases, as shown by Fig.~\ref{fig3}. At scales near the outer correlation length, the PDFs of the increments of the spatial components of ${\bm B}$ and ${\bm E}$ are close to Gaussian, whereas at smaller separations they develop pronounced stretched-exponential tails. This non-Gaussian behavior reflects the formation of intense, localized structures with sharp gradients—such as localized field-line bends, current sheets, and flux ropes \citep{Comisso22}—a characteristic feature of intermittency.

Here we characterize the magnetic-field geometry through the magnetic field-line curvature ${\bm{\kappa}} = {\bm{\hat b}} \cdot {\bm{\nabla}} {\bm{\hat b}}$, with ${\bm{\hat b}} = {\bm B}/B$.\footnote{In practice, we compute the curvature magnitude as $\kappa = |{\bm{\hat b}} \times ({\bm B} \cdot {\bm{\nabla}} {\bm B})|/B^2$, which is algebraically equivalent but more stable in numerical calculations.} Figure~\ref{fig4} shows slices of $B/B_{\rm rms}$ and $\kappa/\kappa_{\rm rms}$ in the $x$–$y$ plane at fixed $z$ for the reference simulation. The maps show a pattern of structures, with regions of strong curvature located in narrow zones that tend to coincide with local minima of the magnetic field. This trend is quantified in the joint probability density of Fig.~\ref{fig4}, which shows that large $\kappa$ values occur preferentially in low-$B$ regions, with a conditional scaling $\langle B \, | \, \kappa \rangle \propto \kappa^{-1/2}$ for $\kappa l_c > 1$. 

\begin{figure}
\begin{center}
   \includegraphics[width=8.65cm]{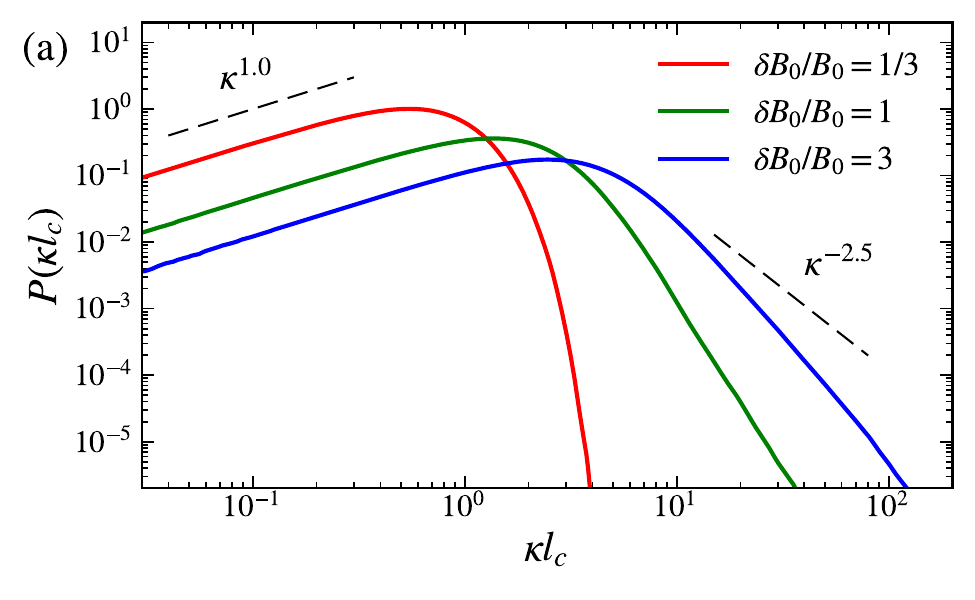}
   \includegraphics[width=8.65cm]{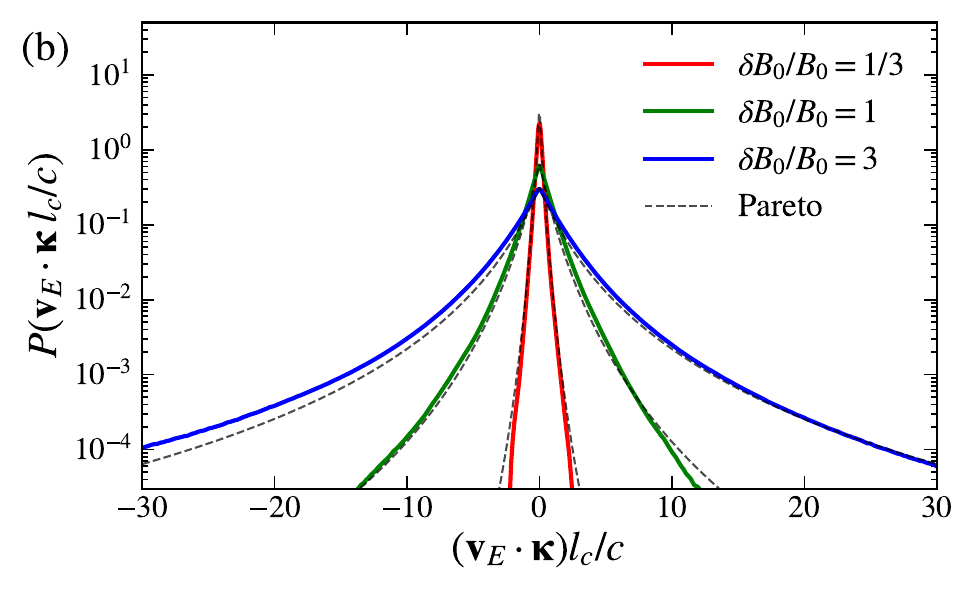}
\end{center}
\vspace{-0.5cm}
   \caption{(a) Probability density functions of the normalized magnetic field-line curvature, $\kappa l_c$, from simulations with different $\delta B_0/B_0$. Power-law slopes $\kappa^{+1}$ and $\kappa^{-2.5}$ are shown for reference. (b) Probability density functions of the normalized magnetic field-line contraction, $({\bm{v}}_E \cdot {\bm{\kappa}})l_c/c$, for the same set of simulations, with symmetric Pareto distribution fits plotted as dashed lines.}
\label{fig5}
\end{figure}

The statistical properties of magnetic field-line curvature depend strongly on the fluctuation-to-mean magnetic field ratio, as illustrated in Fig.~\ref{fig5}. Panel (a) shows the PDFs of $\kappa l_c$ for simulations with different $\delta B_0/B_0$. The distributions are broad, with power-law tails at both low and high curvatures. For reference, power-law slopes of $\kappa^{+1}$ and $\kappa^{-5/2}$ are indicated as asymptotic scalings for low- and high-curvature limits, respectively (see next paragraph). As the mean magnetic field decreases, the peaks of the PDFs shift to higher curvatures, as $\kappa_{\rm rms} \sim l_c^{-1} (1 + B_0^2/\delta B_0^2)^{-1}$, and the distributions extend to $\kappa$ values much larger than the peak, reflecting the strong intermittency of the magnetic field, with localized regions of intense curvature forming within more tangled magnetic configurations.\footnote{We verified that the curvature PDFs remain unchanged when measured in either the simulation frame or the ${\bm{E}} \times {\bm{B}}$ frame.}

The origin of the two power-law regimes can be understood from the decomposition 
\begin{equation}
({\bm{B}} \cdot {\bm{\nabla}}) {\bm{B}} = B \frac{\partial B}{\partial s} {\bm{\hat b}} - f_c {\bm{\hat n}} \, ,
\end{equation}
where $s$ is the curvilinear coordinate along the magnetic field line, $f_c = |{\bm{\hat b}} \times ({\bm B} \cdot {\bm{\nabla}} {\bm B})|$ is the magnitude of the curvature force per unit volume, and ${\bm{\hat b}}$ and ${\bm{\hat n}}$ are unit vectors in the tangential and normal directions. Following \cite{Yang19}, one sees that, since $\kappa = f_c/B^2$, small curvatures ($\kappa \rightarrow 0$) correspond to $f_c \rightarrow 0$ at finite $B$, whereas large curvatures ($\kappa \rightarrow \infty$) correspond to $B \rightarrow 0$ at finite $f_c$. 
In the low-curvature limit, if the two orthogonal components of the curvature force are treated as independent quasi-Gaussian variables (zero mean, variance $\sigma_{f_c}^2$), then $P(f_c) = f_c \sigma_{f_c}^{-2} \exp(-f_c^2/2\sigma_{f_c}^2)$, which reduces to 
\begin{equation}
P(f_c) \sim f_c/\sigma_{f_c}^2 \, , \quad  f_c \rightarrow 0 \, .
\end{equation}
Assuming weak correlations between $f_c$ and $B$, this implies 
\begin{equation}
P(\kappa) \propto \kappa \, , \quad  \kappa \rightarrow 0 \, .
\end{equation}
In the high-curvature limit, with a vanishing guide field, $B_0/\delta B_0 \rightarrow 0$, the three spatial components of the magnetic field can be treated as independent quasi-Gaussian variables (zero mean, variance $\sigma_{B}^2$). In this case $P(B) = \sqrt{2/\pi} B^2 \sigma_B^{-3} \exp(-B^2/2\sigma_B^2)$, so that 
\begin{equation}
P(B) \sim B^2/\sigma_B^{3} \, , \quad  B \rightarrow 0 \, .
\end{equation}
Under the same assumption of weak correlations between $f_c$ and $B$, this leads to 
\begin{equation}
P(\kappa) \propto \kappa^{-5/2} \, , \quad \kappa \rightarrow \infty \, .
\end{equation} 
This scaling has been observed in several studies of non-relativistic turbulence without a mean magnetic field \citep{Yang19,Yuen20,Lemoine23,Kempski23,Kempski25,Kriel25}. In our simulations, we find that the presence of a mean magnetic field progressively steepens the high-$\kappa$ power-law tail from the $-5/2$ scaling, as shown in Fig.~\ref{fig5}(a). When $\delta B_0/B_0 \ll 1$, the power-law behavior is suppressed and the distribution falls off sharply, indicating that large-curvature events, and thus opportunities for strong particle scattering, are markedly reduced.

In Figure~\ref{fig5}(b) we report the PDFs of the magnetic field-line contraction, ${\bm{v}}_E \cdot {\bm{\kappa}}$, normalized by the outer-scale eddy turnover time $l_c/c$, for the same simulations. Here ${\bm{v}}_E = c {\bm{E}} \times {\bm{B}}/B^2$ is the ${\bm{E}} \times {\bm{B}}$ drift velocity. The quantity 
\begin{equation}
{\bm{v}}_E \cdot {\bm{\kappa}} = - \frac{1}{\ell}  \frac{d\ell}{dt}  \, ,
\end{equation}
with $\ell$ the magnetic field-line length, expresses the local rate of field-line shortening or lengthening that accompanies the relaxation or buildup of magnetic tension. Positive values (${\bm{v}}_E \cdot {\bm{\kappa}} > 0$) correspond to tension release, in which field lines shorten and energy is transferred from the field to magnetized particles through curvature-drift motion. Conversely, negative values (${\bm{v}}_E \cdot {\bm{\kappa}} < 0$) correspond to tension buildup, in which field lines are stretched and magnetic energy is accumulated, leading to particle energy loss. In our turbulence simulations, the PDFs of ${\bm{v}}_E \cdot {\bm{\kappa}}$ are approximately symmetric about zero, consistent with Alfv{\'e}nic fluctuations where flow and curvature are out of phase, resulting in negligible net field-line contraction, i.e. $\langle{\bm{v}}_E \cdot {\bm{\kappa}}\rangle l_c/c \sim 0$. This behavior contrasts with large-scale magnetic reconnection geometries \citep{Dahlin17}, which exhibit a net positive ${\bm{v}}_E \cdot {\bm{\kappa}}$ associated with systematic field-line contraction. The near-symmetry of the PDFs in Fig.~\ref{fig5}(b) implies that relaxation and stretching balance on average, so particle energization in turbulence must result from stochastic energy exchanges rather than from a persistent net contraction.

The field-line contraction PDFs display non-Gaussian wings, with heavier tails as the guide field decreases from $B_0 = 3 \delta B_0$ (red) to $B_0 = \delta B_0/3$ (blue). To quantify this behavior, we fit the distributions in Fig.~\ref{fig5}(b) with a symmetric Pareto form, 
\begin{equation}
P({\bm{v}}_E \cdot {\bm{\kappa}}) \propto 
\left(1 + \frac{|{\bm{v}}_E \cdot {\bm{\kappa}}| l_c}{\lambda c} \right)^{-(\alpha+1)} \, ,
\end{equation} 
where $\lambda >0$ is the scale parameter and $\alpha >0$ is the shape parameter. This model reproduces the probability density functions well, with fitted values $\lambda = 2.5, 7, 9$ for $\delta B_0/B_0 = 1/3, 1, 3$, respectively, reflecting the increasing ``typical'' strength of stochastic field-line contraction events as the mean field weakens. The corresponding shape parameters are $\alpha = 8, 5, 3$, indicating that extreme field-line contraction events become more frequent under weaker mean-field conditions. Thus, large-amplitude turbulence not only raises the characteristic amplitude of field-line contraction (through larger $\lambda$) but also enhances intermittency (through smaller $\alpha$), making particle energization more rapid ($\dot{\gamma}/\gamma \propto {\bm{v}}_E \cdot {\bm{\kappa}}$).

\section{Particle Statistics} \label{sec:partstat} 

\begin{figure}
\begin{center}
   \includegraphics[width=8.65cm]{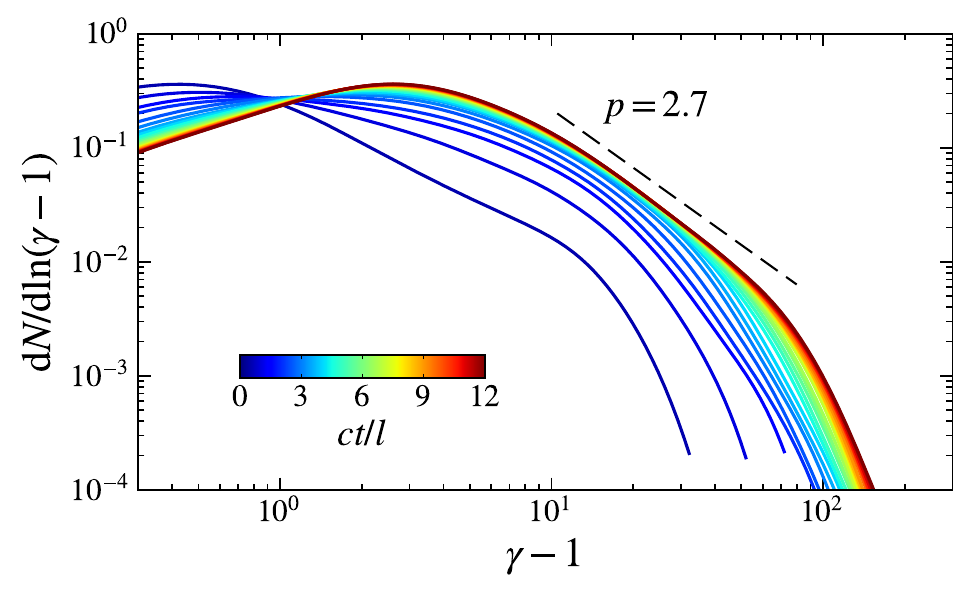}
\end{center}
\vspace{-0.5cm}
   \caption{Time evolution of the proton energy spectrum $dN/d\ln(\gamma-1)$ for the reference simulation ($\delta B_0/B_0=1$). Over time, the particle spectrum develops a high-energy power-law tail with index $p = - d\log N/d\log (\gamma -1) \simeq 2.7$.}
\label{fig6}
\end{figure}

\begin{figure}
\begin{center}
   \includegraphics[width=8.65cm]{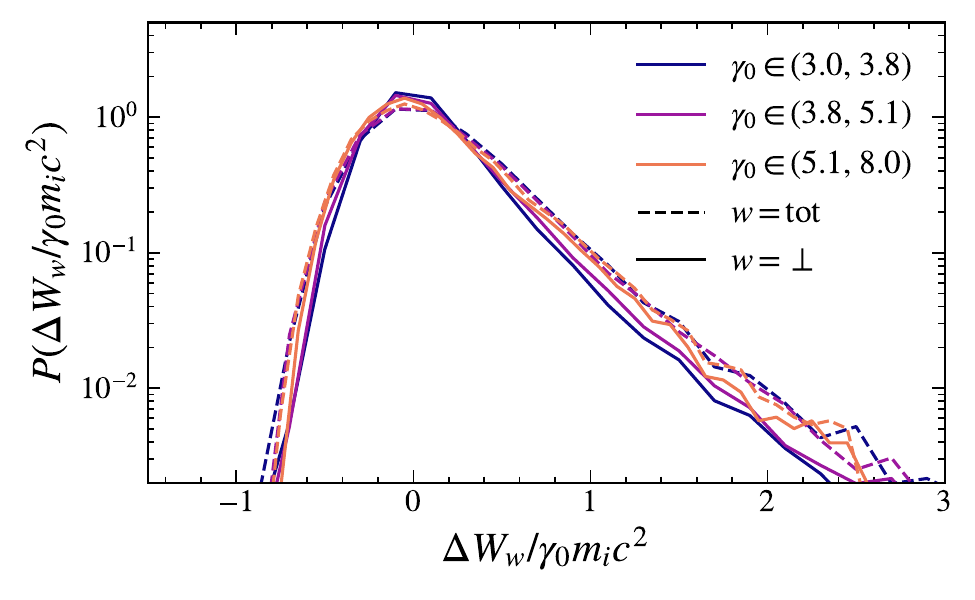}
\end{center}
\vspace{-0.5cm}
    \caption{Probability density functions of $\Delta W_\perp$ and $\Delta W_{\rm tot}$, normalized to the particle energy $\gamma_0 m_i c^2$, accumulated over $\Delta t =0.5\,l_c/c \,$ starting at $t_0=2\,l_c/c$ ($\gamma_0$ is the particle Lorentz factor at $t_0$) in the reference simulation. Different colors correspond to three distinct ranges of $\gamma_0$.}
\label{fig7}
\end{figure}

Figure~\ref{fig6} shows the time evolution of the proton energy spectrum, $dN/d\ln(\gamma-1)$, for the reference simulation. As the system evolves, the spectrum develops a high-energy power-law tail with slope $-2.7$ in $dN/d(\gamma-1)$, consistent with earlier simulations performed with a similar setup \citep{Comisso18,Comisso19}. As shown there, magnetic-field-aligned electric fields associated with magnetic reconnection play a role only in injecting particles from the thermal pool, whereas the subsequent energization, responsible for most of the energy gain, is governed by energy transfer through ${{\bm{E}}_\perp} \cdot {\bm{v}}$ \citep{Comisso18,Comisso19}. To confirm this, we tracked $3 \times 10^5$ randomly selected protons and computed the work done by the component of the electric field perpendicular to the local magnetic field, $\Delta W_\perp = q \int_{t_0}^{t_1} {{{\bm{E}}_\perp}(t) \cdot {\bm{v}}(t) \, dt}$, together with the total electric-field work $\Delta W_{\rm tot}$, both accumulated over a time interval $\Delta t =0.5\,l_c/c$ starting at $t_0=2\,l_c/c$. Figure~\ref{fig7} shows the PDFs of the normalized quantities $\Delta W_\perp/{\gamma_0 m_i c^2}$ and $\Delta W_{\rm tot}/{\gamma_0 m_i c^2}$, where $\gamma_0$ is the particle Lorentz factor at $t_0$. The PDFs are plotted separately for three different $\gamma_0$ bins. The close correspondence between the two distributions indicates that ${{\bm{E}}_\perp}$ accounts for nearly the entire particle energization.

\begin{figure}
\begin{center}
   \includegraphics[width=0.480\textwidth]{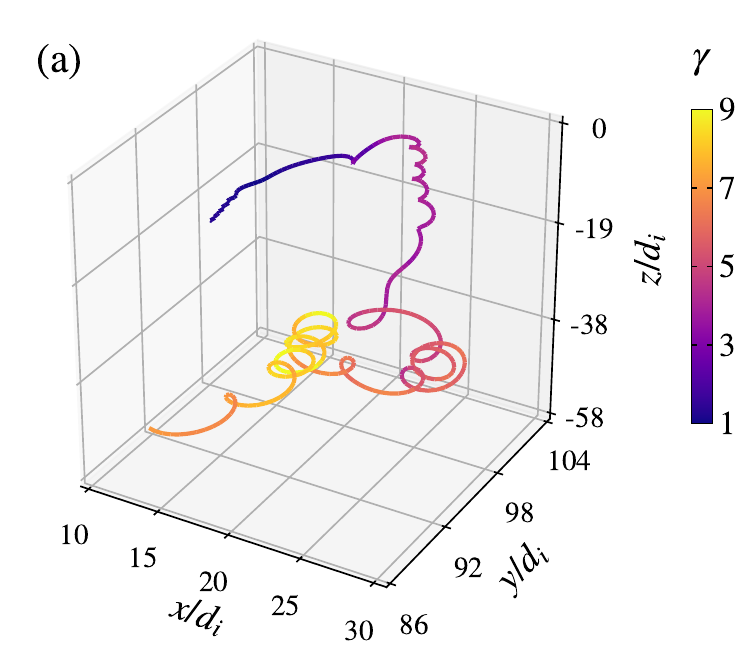}
   \includegraphics[width=0.495\textwidth]{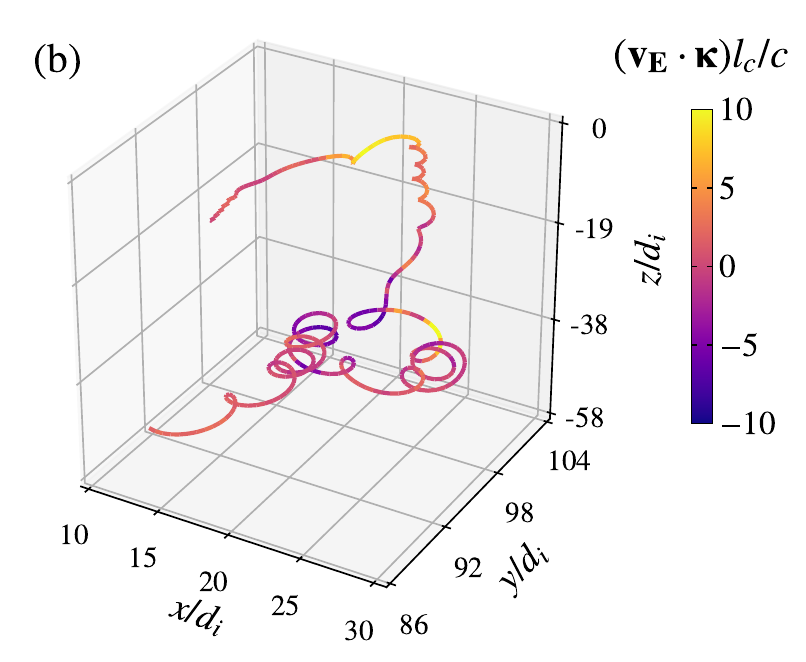}
\end{center}
\vspace{-0.5cm}
   \caption{Example trajectory of a proton from the reference simulation ($\delta B_0/B_0=1$). The top panel shows the trajectory color-coded by the particle Lorentz factor $\gamma$. The bottom panel shows the same trajectory color-coded by the magnetic field-line contraction $({\bm{v}}_E \cdot {\bm{\kappa}}) l_c/c$ at the particle location. The trajectory is plotted over a duration of $4 \,l_c/c$.} 
\label{fig8}
\end{figure}

In what follows, we focus on the ${{\bm{E}}_\perp}$ energization channel, setting aside contributions from ${{\bm{E}}_\parallel}$. To illustrate how a significant fraction of the energization proceeds, Figure~\ref{fig8} presents a representative proton trajectory. The top panel is color-coded by the particle Lorentz factor $\gamma$, while the bottom panel is color-coded by the magnetic field-line contraction $({\bm{v}}_E \cdot {\bm{\kappa}}) l_c/c$ at the particle position. In this example, as the particle moves along its path, episodes of significant energization, where $\gamma$ increases from $\approx 2$ to $\approx 5$, and later from $\approx 5$ to $\approx 7$ in panel (a), occur in association with regions of positive field-line contraction in panel (b), consistent with curvature-driven acceleration. In contrast, intervals of negative field-line contraction coincide with energy losses, such as the decrease in $\gamma$ from $\approx 9$ to $\approx 7$ near the end of this trajectory segment.

To quantify the relative importance of the different energization channels contributing to ${{\bm{E}}_\perp} \cdot {\bm{v}}$, we employ a guiding-center analysis \citep{Northrop1963book}, valid when the electromagnetic fields experienced by a particle vary weakly over a gyroperiod. In this limit, the particle energy $\gamma m_s c^2$ evolves according to \citep{Vandervoort60} 
\begin{equation}\label{eq:denergydt}
\frac{d(\gamma m_s c^{2})}{dt} = q \frac{d \bm{X}}{dt}\cdot \bm{E} + \frac{\mu_M}{\gamma} \frac{\partial}{\partial t} \left(\frac{B}{\gamma_E} \right) \, ,
\end{equation}
where $\mu_M = \gamma^2 v_\perp^2 m_s/2B$ is the magnetic moment, ${\bm{X}}$ is the guiding-center position, and
\begin{equation}
\frac{d \bm{X}}{dt} = {\bm{v}}_{E} + v_\parallel {\bm{\hat b}} + \bm{v}_{\rm{drift}}^{(1)}
\end{equation}
is the guiding-center velocity, with 
\begin{align}
\bm{v}_{\rm drift}^{(1)} 
 &= \frac{\gamma_E^2}{B} \, {\bm{\hat b}} \times \Bigg[ 
      \frac{\gamma m_s c}{q} \Big( v_\parallel {\cal D}({\bm{\hat b}}) 
      + {\cal D}({\bm{v}}_{E}) \Big) \nonumber \\
 &\qquad\qquad
      + \frac{\mu_M}{\gamma q} {\cal G}\!\left(\frac{B}{\gamma_E}\right) 
      + \frac{v_\parallel E_\parallel}{c} {\bm{v}}_{E} \Bigg]
\end{align}
the $1^{\rm th}$-order contribution in the guiding-center expansion. Here $m_s$ is the particle mass for species $s$, $q$ is the electric charge, and $\gamma_E = \left( 1 - E_\perp^2/B^2 \right)^{-1/2}$ is the Lorentz factor associated with ${\bm{v}}_{E}$. The operator ${\cal G}$ is defined as
\begin{equation}
{\cal G} := \frac{{\bm{v}}_{E}}{c} \dfrac{\partial}{\partial t} + c {\bm{\nabla}} \, ,
\end{equation}
while 
\begin{equation}
{\cal D} := \dfrac{\partial }{\partial t} + (v_\parallel {\bm{\hat b}} + {\bm{v}}_{E}) \cdot {\bm{\nabla}} \, .
\end{equation}
The $\bm{E} \cdot \bm{v}_{\rm{drift}}^{(1)}$ term can be expressed in terms of the standard leading contributions, $\bm{E} \cdot \bm{v}_{\rm{drift}}^{(1)} \simeq \bm{E} \cdot ({\bm{v}}_c + {\bm{v}}_p + {\bm{v}}_{\nabla B})$, where 
\begin{equation} \label{eq:v_c}
{\bm{v}}_c = \frac{\gamma_E^2 \gamma m_s c}{q B} v_\parallel \, {\bm{\hat b}} \times  {\cal D}({\bm{\hat b}})
\end{equation}
contains the standard curvature drift velocity, 
\begin{equation}
{\bm{v}}_p = \frac{\gamma_E^2 \gamma m_s c}{q B} \, {\bm{\hat b}} \times {\cal D}({\bm{v}}_{E})
\end{equation}
represents the polarization drift velocity, and 
\begin{equation}
{\bm{v}}_{\nabla B} =  \frac{\gamma_E^2 \mu_M c}{\gamma q B}  \, {\bm{\hat b}} \times {\bm{\nabla}}(B/\gamma_E)
\end{equation} 
is the ${\bm{\nabla}} B$ drift velocity.

\begin{figure}
\begin{center}
   \includegraphics[width=8.65cm]{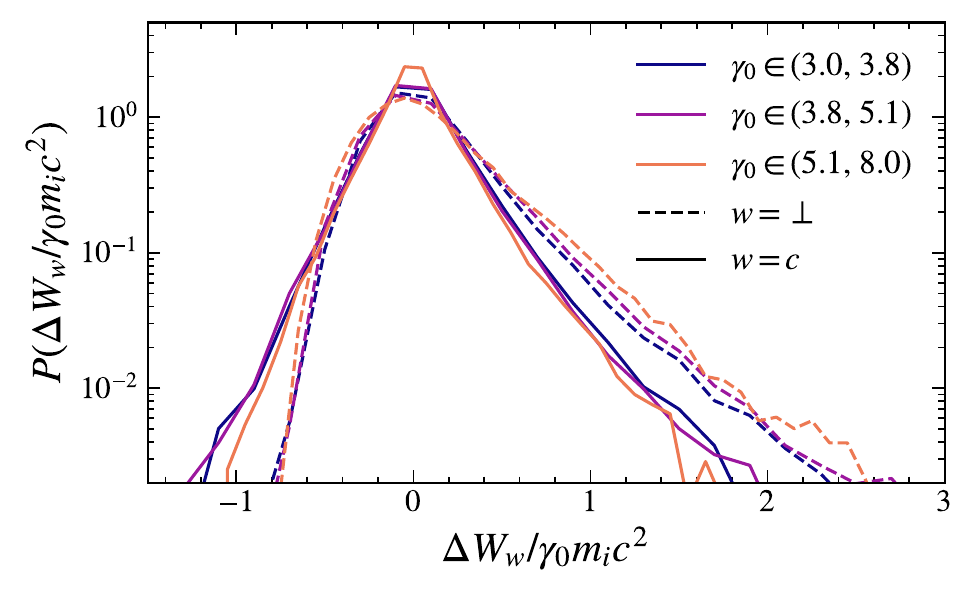}
\end{center}
\vspace{-0.5cm}
   \caption{Probability density functions of $\Delta W_c$ and $\Delta W_\perp$, normalized to the particle energy $\gamma_0 m_i c^2$, accumulated over $\Delta t =0.5\,l_c/c$ starting at $t_0=2\,l_c/c \,$ ($\gamma_0$ is the particle Lorentz factor at $t_0$) in the reference simulation. Different colors correspond to three distinct ranges of $\gamma_0$.}
\label{fig9}
\end{figure}

\begin{figure}
\begin{center}
   \includegraphics[width=8.65cm]{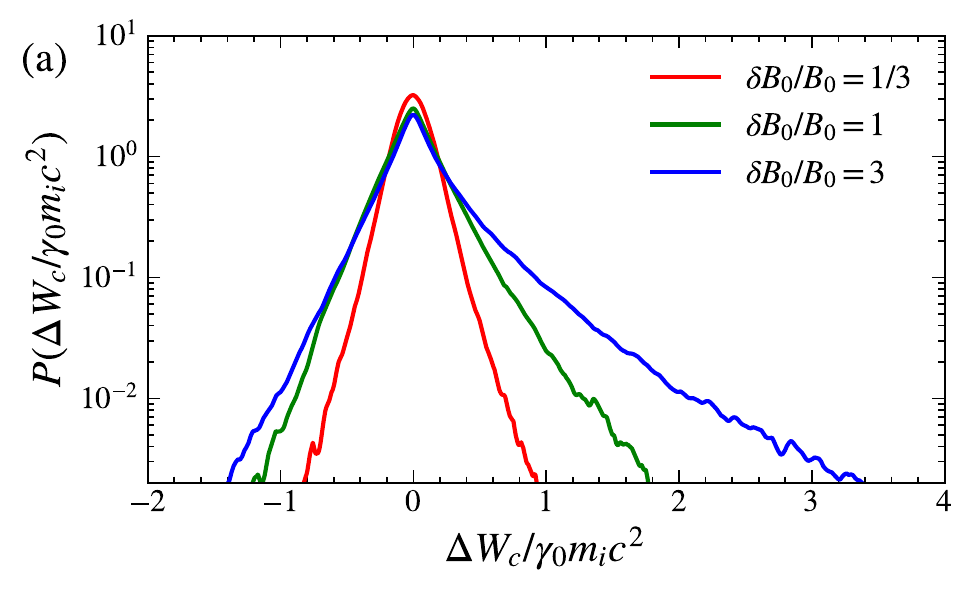}
   \includegraphics[width=8.65cm]{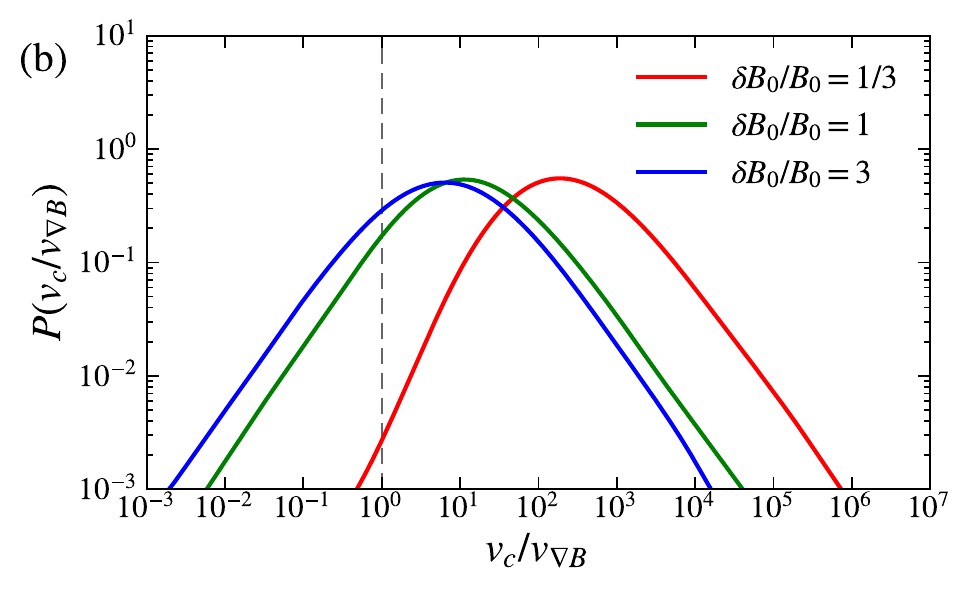}
   \includegraphics[width=8.65cm]{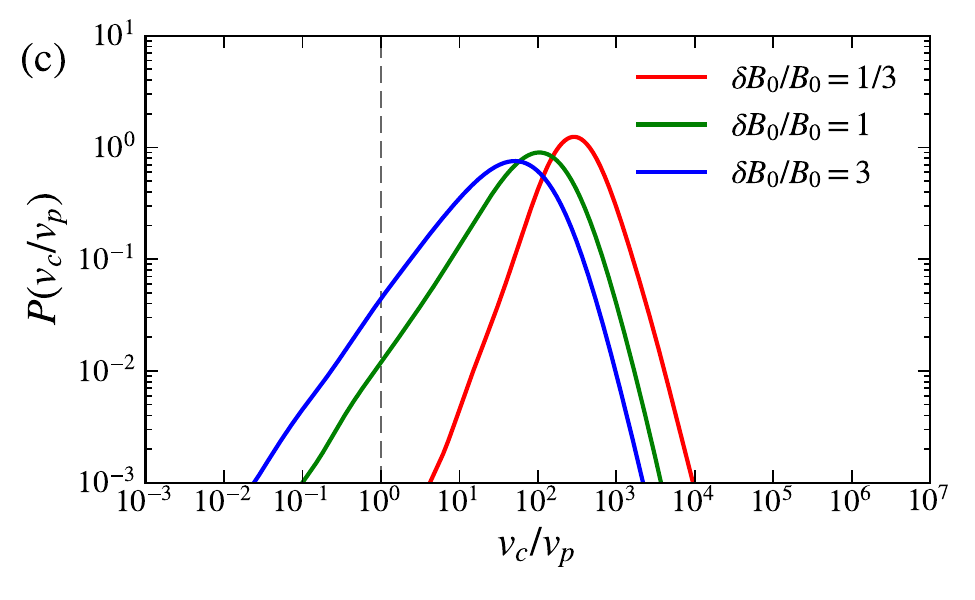}
   \includegraphics[width=8.65cm]{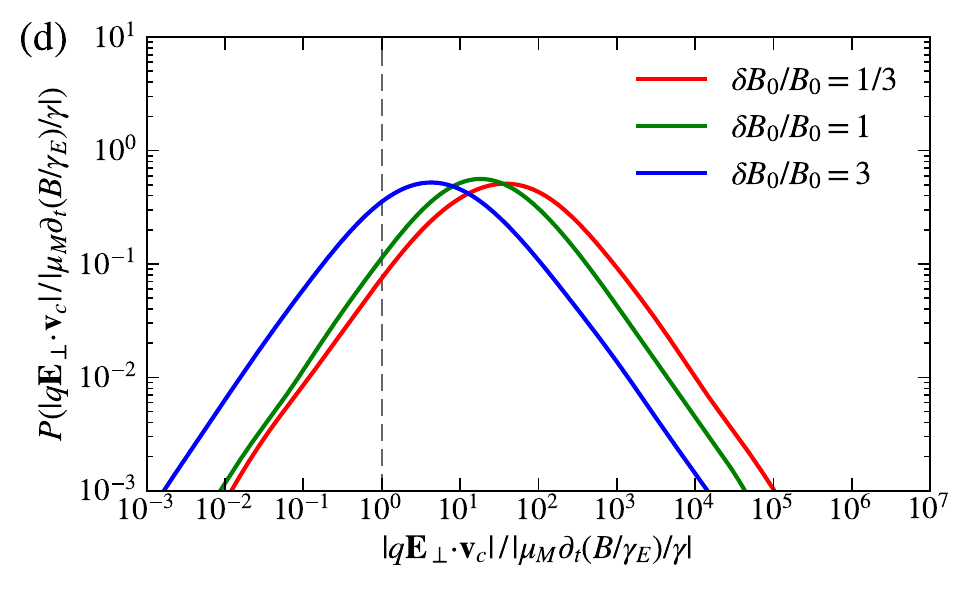}
\end{center}
\vspace{-0.5cm}
   \caption{Dependence of curvature-drift energization and other contributions on $\delta B_0/B_0$. (a) PDFs of $\Delta W_c/{\gamma_0 m_i c^2}$ accumulated over $\Delta t = 0.5 \,l_c/c$ starting at $t_0=2\,l_c/c$, from particles with $\gamma_0 \in (3,8)$. (b–d) PDFs of the ratios $v_c/v_{\nabla B}$, $v_c/v_p$, and $|q{{\bm{E}}_\perp} \cdot {\bm{v}}_c|/|{\mu_M} \partial_t (B/\gamma_E)/{\gamma}|$, respectively, obtained from the tracked particles.}
\label{fig10}
\end{figure}

To connect this formalism with the simulation results, we apply it to the tracked protons in our PIC simulations. In principle, the electromagnetic field should be evaluated at the guiding-center location, but when the particle Larmor radius is sufficiently small compared to the gradient length scale of the magnetic field, the fields at the particle location provide a good approximation. For the particle energies considered here (with $\gamma_0$ a few times the thermal value $\gamma_{{\rm th},i}$) this is a reasonable assumption. We further restrict the analysis to short time segments, $\Delta t = 0.5 \, l_c/c$, since over longer intervals departures from adiabaticity would limit the validity of the guiding-center description.

Focusing on the curvature-drift contribution, Eq.~(\ref{eq:v_c}), we compute the associated work, $\Delta W_c = q \int_{t_0}^{t_1} {{\bm{E}}_\perp}(t)\cdot{\bm v}_c(t) \, dt$ starting at $t_0=2\,l_c/c$, as considered earlier, and compare it with $\Delta W_\perp$. Figure~\ref{fig9} shows that the PDFs of $\Delta W_c/(\gamma_0 m_i c^2)$ make a significant positive contribution to particle energization and closely follow those of $\Delta W_\perp/(\gamma_0 m_i c^2)$. This shows that curvature-drift acceleration accounts for a substantial fraction of the ${\bm E}_\perp \cdot {\bm v}$ energization channel.  

The strength of curvature-drift energization depends on the turbulence level. To assess this dependence, we compare the PDFs of $\Delta W_c/(\gamma_0 m_i c^2)$ for the three values of $\delta B_0/B_0$ explored in our simulations. Here the PDFs are constructed by combining particles across the three energy intervals used in the previous analysis (i.e., $\gamma_0 \in (3,8)$) to improve statistics. As shown in Figure~\ref{fig10}(a), the curvature-drift contribution increases systematically with $\delta B_0/B_0$. In the regime of large-amplitude turbulence ($\delta B_0/B_0 \gtrsim 1$), the PDFs shift markedly toward larger positive values, indicating that curvature drift becomes increasingly effective in accelerating particles.

Other energization channels, such as ${\bm{\nabla}} B$ drift, polarization drift, and betatron energization (the last term in Eq.~(\ref{eq:denergydt})), are also affected by the turbulence level.\footnote{Parallel processes such as magnetic mirroring may also contribute in some settings \citep{Lazarian2023,Vega2023,DasarXiv}, though we do not analyze this mechanism here.} To evaluate their strengths relative to curvature drift, we examine three ratios: (i) $v_c/v_{\nabla B}$ (curvature vs. ${\bm{\nabla}} B$ drift), (ii) $v_c/v_p$ (curvature vs. polarization drift) and (iii) $|q{{\bm{E}}_\perp} \cdot {\bm{v}}_c|/|{\mu_M} \partial_t (B/\gamma_E)/{\gamma}|$ (curvature vs. betatron). The corresponding PDFs, obtained from tracked protons for different values of $\delta B_0/B_0$, are shown in Figs.~\ref{fig10}(b–d). In general, the peaks of these PDFs lie above unity, indicating that curvature drift typically dominates over the other energization channels. As $\delta B_0/B_0$ decreases, the peaks shift to higher values, reflecting that the competing mechanisms weaken more rapidly than curvature drift as the guide field $B_0$ increases. In the limit $\delta B_0/B_0 \ll 1$, on the other hand, parallel energization through ${{\bm{E}}_\parallel} \cdot {\bm{v}}$ \citep{Comisso18,Comisso19} provides the dominant contribution.

\section{Conclusions}\label{sec:conclusions} 

Using fully kinetic PIC simulations, we have investigated the statistical properties of magnetic field-line curvature in highly magnetized turbulence and their impact on particle acceleration. To isolate the effect of the guide-field strength, we varied the fluctuation-to-mean magnetic-field ratio, $\delta B_0/B_0$, and quantified how curvature statistics and acceleration processes change with it.

Regions of strong magnetic field-line curvature correlate with reduced magnetic field strength, and the curvature PDFs display broad power-law wings. Below the peak, the PDFs scale linearly with $\kappa$, while at high $\kappa$ they develop heavy tails for $\delta B_0/B_0 \gtrsim 1$. As the mean field increases, the high-$\kappa$ tail steepens and large-curvature events are strongly suppressed. PDFs of magnetic field-line contraction, ${{\bm v}_E} \cdot {\bm \kappa}$, are well described by symmetric Pareto distributions, with an increased scale parameter and heavier tails resulting from strong intermittency for weaker guide fields. The observed relationships between curvature, magnetic-field strength, and contraction therefore link the geometry of turbulent fields directly to the ability of turbulence to energize particles through the motional electric field, with a strong dependence on the fluctuation-to-mean field ratio.

Within a guiding-center framework, we analyzed the contribution of curvature-drift motion along the motional electric field to particle acceleration. For well-magnetized particles, curvature-drift acceleration constitutes a substantial fraction of the ${\bm E}_\perp \cdot {\bm v}$ channel, and its magnitude increases systematically with $\delta B_0/B_0$. Statistical diagnostics further show that curvature drift typically exceeds the contributions from ${\bm\nabla}B$ drift, polarization drift, and betatron energization across a range of turbulence levels. In regimes dominated by ${\bm E}_\perp$-mediated energy exchange, curvature drift consistently emerges as a principal energization mechanism. 

These results identify curvature-drift acceleration as a key pathway through which magnetized turbulence powers nonthermal particles, clarifying the kinetic basis of turbulent particle acceleration in astrophysical plasmas. A natural next step is to connect curvature-drift statistics with observables such as cosmic-ray spectra, neutrino yields, and variability in high-energy sources.

\vspace{0.5cm}

\section*{Acknowledgments}
We thank James Beattie, Philipp Kempski, and Emanuele Sobacchi for fruitful discussions. We are grateful to Yue Hu for providing the script used to compute the structure functions. This work was supported by the NSF Research Experiences for Undergraduates (REU) program under grant NSF PHY-2447137. L.C. acknowledges support from NSF PHY-2308944 and NASA ATP 80NSSC24K1230. Additional support was provided in part by grant NSF PHY-2309135 to the Kavli Institute for Theoretical Physics (KITP). We acknowledge computing resources from Columbia University's Shared Research Computing Facility project, which is supported by NIH Research Facility Improvement Grant 1G20RR030893-01, and associated funds from the New York State Empire State Development, Division of Science Technology and Innovation (NYSTAR) contract C090171.

\bibliography{curvature_acceleration}{}
\bibliographystyle{aasjournal}

\end{document}